\documentclass[a4paper,10pt]{article}
\usepackage[dvips]{graphicx}
\usepackage{amssymb,amsmath}
\numberwithin{equation}{section}

\begin{document}
\title{\bf{Darboux transformations and exact soliton solutions of integrable coupled spin systems related with  the Manakov system 
}}
\author{ Akbota  Myrzakul\footnote{Email: akbota.myrzakul@gmail.com} \, and    Ratbay Myrzakulov\footnote{Email: rmyrzakulov@gmail.com} \\ \textit{Eurasian International Center for Theoretical Physics and  Department of General } \\ \textit{ $\&$  Theoretical Physics, Eurasian National University, Astana 010008, Kazakhstan}}


\date{}
\maketitle
\begin{abstract}
We construct a Darboux transformation  of a general $su(3)$-valued  spin system called the $\Gamma$-spin system. Using this Darboux transformation we derive a recursive formula for the soliton solutions of this spin system. Then using these results we present explicit expressions for the 1-soliton solution of the coupled $su(2)$-valued spin systems, namely, of the coupled M-LIII equation.
\end{abstract}
\vspace{2cm}
\section{Introduction}
 
Darboux transformations (DT) constitute one of the most fruitful approaches to the construction of soliton solutions of integrable nonlinear equations. DT transform solutions of differential equations into solutions of same class differential equations. There are a key role plays the so-called Darboux matrix: $L$. Here we briefly explain the notion of the Darboux matrix.  For this aim, let us consider the spatial part of the Lax representation 
\begin{eqnarray}
\Re_{x}=U\Re.\label{3.23}
\end{eqnarray}
Consider  the transformation 
\begin{eqnarray}
\Re^{\prime}=L\Re,\label{3.23}
\end{eqnarray}
where Darboux matrix $L$ satisfies the equation
\begin{eqnarray}
L_{x}=U^{\prime}L-LU.\label{3.23}
\end{eqnarray}
Then the $\Re^{\prime}$ obeys the equation
\begin{eqnarray}
\Re_{x}^{\prime}=U^{\prime}\Re^{\prime}.\label{3.23}
\end{eqnarray}
DT originated in the pionering work of Darboux and others on the differential geometry of surfaces. Also we recall that some particular examples of such tranformations were known to Euler and Laplace. In the 1970s, the DT were rediscovered in the soliton theory. In fact, the role of DT is  more important than as mere technical methods for obtaing solutions of some differential equations. 

Integrable spin systems are an important part of integrable systems \cite{Gut}-\cite{akbota6}.  They describe nonlinear dynamics of magnetic systems. 
This article is a continuation of our previous papers devoted to study the integrable coupled spin systems related with the Manakov system and their connections with the integrable  motion of curves and surfaces (see e.g.  \cite{akbota1}-\cite{akbota6} and references therein). In this paper we consider the DT for integrable coupled spin systems namely for the coupled M-LIII equation and for the $\Gamma$-spin system \cite{akbota1}-\cite{akbota6}. We recall that these   spin systems are equivalent to the one and same equation namely to the Manakov system. These three equations - the coupled M-LIII equation, the $\Gamma$-spin system and the Manakov system are integrable by the inverse scattering method. The coupled M-LIII equation is one of natural candidates to be  integrable multilayer  generalizations of the classical continious Heisenberg ferromagnetic equation.

The structure of the paper is  as follows.  In the next two sections we gather a number of definitions and basic results on the Manakov system and the M-LIII equation. These are needed to construct the DT and the exact soliton solutions of these two integrable systems. Section 2 provides a basic review of the Manakov system. In Section 3, we recall key results on  the coupled M-LIII equation. In particular we give the  LR for the coupled M-LIII  equation.   In Section 4, we establish the gauge equivalence between the coupled  M-LIII equation and the Manakov system.  Sections 5 and 6 are devoted to  the DT for the $\Gamma$-spin system and for the coupled M-LIII equation.  Finally, in  Sec. 7, we present our conclusions. 

\section{The Manakov system. Preliminaries}
In this section we recall some general facts about the Manakov system. The Manakov system  has many physical significant applications such as the modeling crossing sea waves and for propagation in elliptically birefringent optical fibers.  
\subsection{The equation}
The Manakov system is the particular case of the vector nonlinear Schrodinger equation and has the form
\begin{eqnarray}
iq_{1t}+q_{1xx}+2(|q_{1}|^{2}+|q_{2}|^{2})q_{1}&=&0,\label{3.21}\\
 iq_{2t}+q_{2xx}+2(|q_{1}|^{2}+|q_{2}|^{2})q_{2}&=&0.\label{3.23}
\end{eqnarray}

\subsection{Lax representation}

The Manakov  equation is integrable by the inverse scattering method. Its Lax representation reads as 
\begin{eqnarray}
\Phi_{x}&=&U\Phi,\label{2.1}\\
\Phi_{t}&=&V\Phi, \label{2.2} 
\end{eqnarray}
where $\Phi=(\phi_{1},\phi_{2}, \phi_{3})$ and 
\begin{eqnarray}
U=-i\lambda \Sigma+U_{0}, \quad V=-2i\lambda^{2}\Sigma+2\lambda U_{0}+V_{0}. \label{2.2} 
\end{eqnarray}
Here
\begin{eqnarray}
\Sigma =
\left ( \begin{array}{ccc}
1   & 0     & 0 \\
0 & -1    & 0  \\
0   & 0 & -1
\end{array} \right),\, U_{0} =
\left ( \begin{array}{ccc}
0       & q_{1}  & q_{2} \\
-\bar{q}_{1} & 0      & 0 \\
-\bar{q}_{2}  & 0 & 0
\end{array} \right), \,   V_{0} =
i\left ( \begin{array}{ccc}
|q_{1}|^{2}+|q_{2}|^{2}      & q_{1x}  & q_{2x} \\
\bar{q}_{1x} & - |q_{1}|^{2}     & -\bar{q}_{1}q_{2} \\
\bar{q}_{2x}  & -\bar{q}_{2}q_{1} & -|q_{2}|^{2}
\end{array}\right).\label{2.2} 
\end{eqnarray}
\section{The coupled M-LIII equation. Preliminaries}
 The coupled M-LIII equation describe nonlinear dynamics of the coupled (two-layer) magnetic systems. It is integrable by the inverse scattering method.  The coupled M-LIII equation plays important role in differential geometry of curves and surfaces. It related with the fact that this coupled equation induced the integrable class of two interacting curves and surfaces.  In this section we give some main facts on this equation.
 
  \subsection{The equation}
 
 Consider two spin vectors ${\bf A}=(A_{1}, A_{2}, A_{3})$ and  ${\bf B}=(B_{1}, B_{2}, B_{3})$, where ${\bf A}^{2}={\bf B}^{2}=1$. Then  the  coupled  M-LIII equation reads as
\begin{eqnarray}
iA_{t}+\frac{1}{2}[A,A_{xx}]+iu_{1}A_{x}+F&=&0,\\
iB_{t}+\frac{1}{2}[B,B_{xx}]+iu_{2}B_{x}+E&=&0, \label{2.2} 
\end{eqnarray}
where $u_{k}$
are real functions, $F$ and $E$  are matrix functions, 
\begin{eqnarray}
A=\begin{pmatrix} A_{3}&A^{-}\\ 
A^{+}&-A_{3}\end{pmatrix},\quad 
B=\begin{pmatrix} B_{3}&B^{-}\\ 
B^{+}&-B_{3}\end{pmatrix}, \quad F=\begin{pmatrix} F_{3}&F^{-}\\ 
F^{+}&-F_{3}\end{pmatrix}, \quad
E=\begin{pmatrix} E_{3}&E^{-}\\ 
E^{+}&-E_{3}\end{pmatrix},  \label{2.2} 
\end{eqnarray}
with  $A^{\pm}=A_{1}\pm iA_{2}, \quad B^{\pm}=B_{1}\pm iB_{2}, \quad A^{2}=B^{2}=I, \quad F^{\pm}=F_{1}\pm iF_{2},\quad E^{\pm}=E_{1}\pm iE_{2}$. Here  $A\in su(2), \quad B\in su(2)$.  
In this paper, we assume that $F$ and $E$ have the form
\begin{eqnarray}
F=v_{1}[\sigma_{3},A],\quad E=v_{2}[\sigma_{3},B], \label{2.2} 
\end{eqnarray}
where  $v_{j}$ are some real functions (potentials). 
Then the  coupled M-LIII equation takes the form
\begin{eqnarray}
iA_{t}+\frac{1}{2}[A,A_{xx}]+iu_{1}A_{x}+v_{1}[\sigma_{3},A]&=&0,\\
iB_{t}+\frac{1}{2}[B,B_{xx}]+iu_{2}B_{x}+v_{2}[\sigma_{3},B]&=&0, \label{2.2} 
\end{eqnarray}
where $u_{j}$ and $v_{j}$ are coupling potentials and have the forms
\begin{eqnarray}
u_{1}&=&\frac{2i}{\Delta_{1}}\left({\bar q}_{2}g_{1}{\bar g}_{3}-q_{2}{\bar g}_{1}g_{3}\right),\label{3.666}\\
v_{1}&=&-\frac{|q_{2}|^{2}\Delta}{\Delta_{1}},\label{2.16}\\
u_{2}&=&\frac{2i}{\Delta_{2}}\left({\bar q}_{1}g_{1}{\bar g}_{2}-q_{1}{\bar g}_{1}g_{2}\right),\label{3.666}\\
v_{2}&=&-\frac{|q_{1}|^{2}\Delta}{\Delta_{2}}.\label{2.16}
\end{eqnarray} 
Here 
\begin{eqnarray}
\Delta_{1}&=&|g_{1}|^{2}+|g_{2}|^{2},\label{3.666}\\
\Delta_{2}&=&|g_{1}|^{2}+|g_{3}|^{2},\label{3.666}\\
\Delta&=&|g_{1}|^{2}+|g_{2}|^{2}+|g_{3}|^{2}.\label{3.666}
\end{eqnarray} In  elements,  the coupled M-LIII equation (3.5)-(3.6) reads as
\begin{eqnarray}
iA_{t}^{+}+(A^{+}A_{3xx}-A^{+}_{xx}A_{3})+iu_{1}A^{+}_{x}-2v_{1}A^{+}&=&0,\\
iA_{t}^{-}-(A^{-}A_{3xx}-A^{-}_{xx}A_{3})+iu_{1}A_{x}^{-}+2v_{1}A^{-}&=&0,\\
iA_{3t}+\frac{1}{2}(A^{-}A^{+}_{xx}-A^{-}_{xx}A^{+})+iu_{1}A_{3x}&=&0,\\
iB_{t}^{+}+(B^{+}B_{3xx}-B^{+}_{xx}B_{3})+iu_{2}B_{x}^{+}+2v_{2}B^{+}&=&0, \label{2.2} \\
iB_{t}^{-}-(B^{-}B_{3xx}-B^{-}_{xx}B_{3})+iu_{2}B_{x}^{-}+2v_{2}B^{-}&=&0, \label{2.2} \\
iB_{3t}+\frac{1}{2}(B^{-}B^{+}_{xx}-B^{-}_{xx}B^{+})+iu_{2}B_{3x}&=&0. \label{2.2} 
\end{eqnarray}

\subsection{Lax representation}
The coupled M-LIII  equation is integrable by the inverse scattering method. It admits the Lax representation of the  form 
\begin{eqnarray}
Y_{x}&=&(-i\lambda A+S_{1})Y+H_{1}\label{2.1}\\
Y_{t}&=&-2i\lambda^{2}AY+2\lambda N_{1}Y+X_{1}Y+J_{1},\label{2.1} 
\end{eqnarray}
\begin{eqnarray}
Z_{x}&=&(-i\lambda B+S_{2})Z+H_{2}\label{2.1}\\
Z_{t}&=&-2i\lambda^{2}BZ+2\lambda N_{2}Z+X_{2}Z+J_{1},\label{2.1} 
\end{eqnarray}
where $Y=(\psi_{1}, \psi_{2})^{T}$ and  $Z=(\psi_{4}, \psi_{3})^{T}$, $S_{j}$ and $H_{j}$ are  matrix finctions.
The compatibility conditions $Y_{xt}=Y_{tx}$ and $Z_{xt}=Z_{tx}$ give the coupled M-LIII equation (3.5)-(3.6).
\section{Gauge equivalence between the coupled M-LIII equation and the Manakov system}
 Let $Y=(\psi_{1}, \psi_{2})^{T}$ and  $Z=(\psi_{4}, \psi_{3})^{T}$ are the solutions of the set 3.20)-(3.21). Now let us  introduce the new functions $\phi_{j}$ as
 \begin{eqnarray}
\phi_{1}=g_{1}\psi_{1}+{\bar g}_{2}\psi_{2}, \quad \phi_{2}=g_{2}\psi_{1}-{\bar g}_{1}\psi_{2}, \quad \phi_{3}=g_{1}\psi_{1}+{\bar g}_{2}\psi_{3}.
\end{eqnarray}
 
 The straight calculations show that these new functions $\phi_{j}$ satisfy the set of equations (2.3)-(2.4). That is they give the Lax representation of the Manakov system. This result proves the gauge equivalence between the coupled M-LIII equation (3.5)-(3.6) and the Manakov system (2.1)-(2.2) \cite{akbota1}-\cite{akbota6}. 
\section{Gauge equivalence between the $\Gamma$-spin system and the Manakov system}
 We recall  that there is the another spin system which is the gauge equivalent to the Manakov system (2.1)-(2.2). This spin system called by us as,  the $\Gamma$-spin system, reads as \cite{Kostov}
 \begin{eqnarray}
i\Gamma_{t}+\frac{1}{2}[\Gamma, \Gamma_{xx}]=0. \label{2.2} 
\end{eqnarray}
It is integrable and admits the following  Lax representation (see e.g. \cite{Kostov} and references therein)
\begin{eqnarray}
\Psi_{x}&=&U^{\prime}\Psi,\label{2.1}\\
\Psi_{t}&=&V^{\prime}\Psi, \label{2.2} 
\end{eqnarray}
where
\begin{eqnarray}
U^{\prime}=-i\lambda\Gamma, \quad V^{\prime}=-2i\lambda^{2}\Gamma+\frac{1}{2}\lambda[\Gamma, \Gamma_{x}]. \label{2.2} 
\end{eqnarray}
Here
\begin{eqnarray}
\Gamma=g^{-1}\Sigma g, \quad \Gamma^{2}=I \label{2.2} 
\end{eqnarray}
and
\begin{eqnarray}
\Gamma =
\left ( \begin{array}{ccc}
\Gamma_{11}  & \Gamma_{12}    &\Gamma_{13} \\
\Gamma_{21} &   \Gamma_{22} &\Gamma_{23} \\
\Gamma_{31}   & \Gamma_{32} & \Gamma_{33}
\end{array} \right)\in su(3).\label{2.2} 
\end{eqnarray}

 \section{Darboux transformation and exact solutions of the $\Gamma$-spin system}
 \subsection{Darboux transformation of the $\Gamma$-spin system}
 In this section, we construct  the DT for the equation (5.1). To do this, let us  consider the following transformation of solutions of the  equations (5.2)-(5.3)
\begin{eqnarray}
\Psi^{\prime}=L\Psi, \label{3.1}
\end{eqnarray}
where 
\begin{eqnarray}
L=\lambda N-I. \label{3.2}
\end{eqnarray}
We require  that $\Phi^{\prime}$ satisfies the same Lax representation as (5.2)-(5.3) so that
\begin{eqnarray}
\Phi'_{x} &=& U'\Phi',\label{3.3}\\
\Phi'_{t} &=& V'\Phi', \label{3.4}
\end{eqnarray}
where $U^{\prime}-V^{\prime}$ depend on $\Gamma^{\prime}$  as $U-V$ on $\Gamma$. The matrix  $L$ obeys the  following equations
\begin{eqnarray}
L_{x}+LU &=& U'L,\label{3.6} \\
L_{t}+LV &=& V'L. \label{3.7}
\end{eqnarray}
 These equations yield  the following equations for $N$ 
   \begin{eqnarray}
   N_{x}&=&i\Gamma^{'}-i\Gamma, \label{3.10}\\
   N_{t}&=&-\Gamma'\Gamma'_x+\Gamma\Gamma_x \label{3.15}
\end{eqnarray}
and 
\begin{eqnarray}
   \Gamma^{'}=N\Gamma N^{-1}. \label{3.20}
\end{eqnarray}
Also we have  the following useful second form of the DT for $S$:
\begin{eqnarray}
   \Gamma^{'}=\Gamma-iN_{x}.  \label{3.24}
\end{eqnarray}

 Let us consider the following set of equations
\begin{eqnarray}
H_{x} &=& -i\Gamma H\Lambda, \label{3.50}\\
H_{t} &=& -2i\Gamma H\Lambda^2+\Gamma\Gamma_xH\Lambda, \label{3.51}
\end{eqnarray}
where 
\begin{eqnarray}
\Lambda=\begin{pmatrix} \lambda_{1}&0&0\\0&\lambda_{2}&0\\0&0&\lambda_{3}\end{pmatrix},   \label{3.52}
\end{eqnarray}
$det$ $H\neq0$ and  $\lambda_{k}$  are complex constants. We now assume that  the matrix $N$ can be written as:
\begin{eqnarray}
N=H\Lambda^{-1} H^{-1}=\begin{pmatrix} n_{11}&n_{12}& n_{13}\\ 
  n_{21}& n_{22}&n_{23}\\ 
n_{31}&n_{32}&n_{33} \end{pmatrix}. \label{3.47}
\end{eqnarray}
The inverse matrix we write as
\begin{eqnarray}
N^{-1}=H\Lambda H^{-1}=\frac{1}{n}\begin{pmatrix} m_{11}&m_{12}& m_{13}\\ 
  m_{21}& m_{22}&m_{23}\\ 
m_{31}&m_{32}&m_{33} \end{pmatrix} \label{3.47}
\end{eqnarray}
or
\begin{eqnarray}
N^{-1}=\frac{1}{n}\begin{pmatrix} n_{22}n_{33}-n_{23}n_{32}&-(n_{12}n_{33}-n_{13}n_{32})& n_{12}n_{23}-n_{13}n_{22}\\ 
 -(n_{21}n_{33}-n_{23}n_{31})& n_{11}n_{33}-n_{31}n_{13}&-(n_{11}n_{23}-n_{21}n_{13})\\ 
n_{32}n_{21}-n_{31}n_{22}&-(n_{11}n_{32}-n_{31}n_{12})&n_{11}n_{22}-n_{21}n_{12} \end{pmatrix}. \label{3.47}
\end{eqnarray}
where $n=det{N}$ and has the form
\begin{eqnarray}
n=n_{11}n_{22}n_{33}+n_{12}n_{23}n_{31}+n_{13}n_{32}n_{21}-n_{31}n_{22}n_{13}-n_{12}n_{21}n_{33}-n_{11}n_{23}n_{32}. \label{3.47}
\end{eqnarray}
  From these equations follow that $N$ obeys the equations
  \begin{eqnarray}
N_{x} &=& iN\Gamma N^{-1}-i\Gamma, \label{3.53}\\
N_{t} &=& \Gamma\Gamma_x-N\Gamma\Gamma_xN^{-1}, \label{3.55}
\end{eqnarray}
which are equivalent to Eqs.(6.7)-(6.8) as we expected.  The $\Gamma$ and matrix solutions of the system (5.2)-(5.3) obey the condition
\begin{eqnarray}
\Phi^{\dagger}=\Phi^{-1}, \quad \Gamma^{\dagger}=\Gamma, \label{3.62}
\end{eqnarray} 
which follow from the equations
\begin{eqnarray}
\Phi^{\dagger}_{x}=i\lambda \Phi^{\dagger}\Gamma^{\dagger}, \quad (\Phi^{-1})_{x}=i\lambda \Phi^{-1}\Gamma^{-1}. \label{3.63}
\end{eqnarray}
Here $\dagger$ denote an Hermitian conjugate. After some calculations we came to the formulas 
\begin{eqnarray}
\lambda_{2}=\lambda_{3}=\lambda^{*}_{1}, \quad
 H=\begin{pmatrix} \psi_{1}(\lambda_{1};t,x,y)&\psi^{*}_{2}(\lambda_{1};t,x,y)&\psi^{*}_{3}(\lambda_{1};t,x,y)\\  \psi_{2}(\lambda_{1};t,x,y)&-\psi^{*}_{1}(\lambda_{1};t,x,y)&0\\ 
\psi_{3}(\lambda_{1};t,x,y)&0&-\psi^{*}_{1}(\lambda_{1};t,x,y)\\ \end{pmatrix}, \label{3.64}
\end{eqnarray}
\begin{eqnarray}
H^{-1}=\frac{1}{\square{\bar \psi}_{1}}\begin{pmatrix} {\bar \psi}^{2}_{1}&{\bar \psi}_{1}{\bar \psi}_{2}&{\bar \psi}_{1}{\bar \psi}_{3}\\  {\bar \psi}_{1}\psi_{2}&-(|{\bar \psi}_{1}|^{2}+|\psi_{2}|^{2})& \psi_{2}{\bar \psi}_{3}\\ 
{\bar \psi}_{1}\psi_{3}&{\bar \psi}_{2} \psi_{3}&-(|{\bar \psi}_{1}|^{2}+|\psi_{2}|^{2}\\ \end{pmatrix}, \label{3.64}
\end{eqnarray}
where 
\begin{eqnarray}
\square &=&|\psi_{1}|^2+|\psi_{2}|^2+|\psi_{3}|^2. \label{3.66}
\end{eqnarray}
So finally for the matrix $N$ we get the following expression
\begin{eqnarray}
 N=\begin{pmatrix} n_{11}&n_{12}& n_{13}\\ 
  n_{21}& n_{22}&n_{23}\\ 
n_{31}&n_{32}&n_{33} \end{pmatrix}=\frac{1}{\square}\begin{pmatrix} n_{11}\square&n_{12}\square& n_{13}\square\\ 
  {\bar \psi}_{1}\psi_{2}\epsilon_{12}&\frac{|\psi_{2}|^{2}}{\lambda_{1}}+\frac{|{\bar \psi}_{1}|^{2}+|\psi_{3}|^{2}}{\lambda_{2}}& \psi_{2}{\bar \psi}_{3}\epsilon_{12}\\ 
{\bar \psi}_{1}\psi_{3}\epsilon_{13}&{\bar \psi}_{2} \psi_{3}\epsilon_{13}&\frac{|\psi_{3}|^{2}}{\lambda_{1}}+\frac{|\psi_{1}|^{2}+|\psi_{2}|^{2}}{\lambda_{3}}\\ \end{pmatrix},   \label{3.67}
\end{eqnarray}
where $\epsilon_{ij}=\lambda_{i}^{-1}-\lambda_{j}^{-1}$, 
\begin{eqnarray}
n_{11}\square&=& \frac{| \psi_{1}|^{2}}{\lambda_{1}}+\frac{| \psi_{2}|^{2}}{\lambda_{2}}+\frac{| \psi_{3}|^{2}}{\lambda_{3}},   \\
n_{12}\square&=& \frac{ \psi_{1}{\bar \psi}_{2}}{\lambda_{1}}-\frac{ \psi_{1}{\bar \psi}_{2}}{\lambda_{2}}+\frac{ \psi_{2}|\psi_{3}|^{2}}{{\bar \psi}_{1}}(\lambda_{3}^{-1}-\lambda_{2}^{-1}),   \\
n_{13}\square&=& \frac{ \psi_{1}{\bar \psi}_{3}}{\lambda_{1}}-\frac{ \psi_{1}{\bar \psi}_{3}}{\lambda_{3}}+\frac{ |\psi_{2}|^{2}{\bar \psi}_{3}}{{\bar \psi}_{1}}(\lambda_{2}^{-1}-\lambda_{3}^{-1}).
\end{eqnarray}
Hence we can write the DT in terms of the eigenfunctions of the Lax representations (5.2)-(5.3) as
\begin{eqnarray}
\Gamma^{[1]}=\frac{1}{n}\begin{pmatrix} n_{11}m_{11}-n_{12}m_{21}-n_{13}m_{31}&n_{11}m_{12}-n_{12}m_{22}-n_{13}m_{32}& n_{11}m_{13}-n_{12}m_{23}-n_{13}m_{33}\\ 
n_{21}m_{11}-n_{22}m_{21}-n_{23}m_{31}&n_{21}m_{12}-n_{22}m_{22}-n_{23}m_{32}&n_{21}m_{13}-n_{22}m_{23}-n_{23}m_{33}\\ 
n_{31}m_{11}-n_{32}m_{21}-n_{33}m_{31}&n_{31}m_{12}-n_{32}m_{22}-n_{33}m_{32}&n_{31}m_{13}-n_{32}m_{23}-n_{33}m_{33} \end{pmatrix}  
\label{3.68}
\end{eqnarray}

\subsection{1-Soliton solutions for $\Gamma$ - spin system}

To construct the 1-soliton solution of the $\Gamma$-spin system (5.1), now we consider a seed solution 
 \begin{eqnarray}
    \Gamma^{[0]}=\Sigma. \label{4.17}
\end{eqnarray}
In our case the eigenfunctions are given by
\begin{eqnarray}
\psi_{1}=e^{-i\lambda x-2i\lambda^2t+i\delta_{1}}=e^{-\theta+i\delta_{1}},\quad \psi_{2}=e^{\theta+i\delta_{2}}, \quad \psi_{3}=e^{\theta+i\delta_{3}},\label{4.33}\end{eqnarray}
where  $\delta_i$ are complex constants and 
 \begin{eqnarray}
 \theta=\theta_{1}+i\theta_{2}=-i\lambda_{1} x-2i\lambda_{1}^{2}t. 
 \end{eqnarray}
 Then we get
\begin{eqnarray}
\Gamma^{[1]}=
\left ( \begin{array}{ccc}
\Gamma_{11}^{[1]}  & \Gamma_{12}^{[1]}    &\Gamma_{13}^{[1]} \\
\Gamma_{21}^{[1]} &   \Gamma_{22}^{[1]} &\Gamma_{23}^{[1]} \\
\Gamma_{31}^{[1]}   & \Gamma_{32}^{[1]} & \Gamma_{33}^{[1]}
\end{array} \right),\label{2.2} 
\end{eqnarray}
where $n_{ij}$ and $m_{ij}$ are given by the equations (6.15)-(6.16) and (6.25).
\section{The  1-soliton solutions of  the  coupled M-LIII equation}
Let us now we present the formulas for the 1-soliton solution of the coupled M-LIII equation (3.5)-(3.6). Its  seed solution we write as 
\begin{eqnarray}
A^{[0]} =\sigma_{3}, \quad B^{[0]} =\sigma_{3}.\label{2.2} 
\end{eqnarray}
To find the 1-soliton solution of the coupled M-LIII equation here we use  the inverse M-transformation \cite{akbota1}-\cite{akbota6}. The inverse  M-transformation allows us to fine solutions of  the coupled M-LIII equation (3.5)-(3.6),  if we know the solutions of the $\Gamma$-spin system (5.1). So the 1-soliton solution of the coupled M-LIII equation has the form
\begin{eqnarray}
A^{[1]} &=&\frac{1}{1-\Gamma_{33}^{[1]}}
\left ( \begin{array}{cc}
\Gamma_{11}^{[1]}-\Gamma_{22}^{[1]}   & 2\Gamma_{12}^{[1]} \\
2\Gamma_{21}^{[1]} &   \Gamma_{22}^{[1]}-\Gamma_{11}^{[1]}\end{array}\right),\label{2.2} \\
B^{[1]}&=&\frac{1}{1-\Gamma_{22}^{[1]}}
\left ( \begin{array}{cc}
\Gamma_{11}^{[1]}-\Gamma_{33}^{[1]}   & 2\Gamma_{13}^{[1]} \\
2\Gamma_{31}^{[1]} &   \Gamma_{33}^{[1]}-\Gamma_{11}^{[1]}\end{array}\right),\label{2.2} 
\end{eqnarray}
 where $\Gamma_{ij}^{[1]}$ are given by the formulas (6.33) and (6.29).

 \section{Conclusion}
 In this paper we have presented the DT for the $\Gamma$-spin system which is the integrable $su(3)$-valued spin system.  In particular, we have given the explicit formula for its  1-soliton solution. Then  we  have shown how construct soliton solutions of the coupled M-LIII equation for the two coupled $su(2)$ - valued  spin systems.  For this purpose we have used the  DT formulas of the $\Gamma$-spin system. 
 Also the Lax representation of the coupled M-LIII equation is presented. Using this Lax representation, the gauge equivalence between the coupled M-LIII equation and the Manakov system is established. The results obtained in this paper will be useful in the study of nonlinear dynamics of multi-layer magnetic systems. Also they will be useful in differential geometry of curves and surfaces to find their integrable deformations of interacting curves and surfaces.

 \end{document}